\documentclass[12pt,reqno]{amsart}
\usepackage[foot]{amsaddr}
\usepackage{xr}
\usepackage{amsthm}
\usepackage[mathscr]{eucal}
\usepackage{amsfonts}
\usepackage{amsmath}
\usepackage{amssymb}
\usepackage{graphicx}

\usepackage{pstricks}
\usepackage{fullpage}
\usepackage{hyperref}
\usepackage{booktabs}
\usepackage{colortbl}

\usepackage{enumitem}

\usepackage{caption}
\usepackage{subcaption}
\usepackage{setspace}
\usepackage{tabularx}
\usepackage{natbib}
\usepackage{booktabs}

\newcommand{\mb}[1]{\mathbb{#1}}

\newcommand{\bs}[1]{\boldsymbol{#1}}

\newcommand{\tn}[1]{\textnormal{#1}}

\newcommand{\ind}{{1\hspace{-2.5pt}\tn{l}}}

\newcommand{\R}{\mb{R}}

\def\beqn{\begin{eqnarray*}}
\def\eeqn{\end{eqnarray*}}
\def\beq{\begin{eqnarray}}
\def\eeq{\end{eqnarray}}
\def\bm#1{\mbox{\boldmath{$#1$}}}

\def\nullvec {\mathbf{0}}

 \DeclareMathOperator{\Cor}{\mathbb{C}\mbox{or}}

    \def \mC {\text{\boldmath$C$}}
    \def \mD {\text{\boldmath$D$}}

    \def \mI {\text{\boldmath$I$}}

    \def \mQ {\text{\boldmath$Q$}}
\def \rvec {\text{\boldmath$r$}}

\def \vvec {\text{\boldmath$v$}}    
    
    \def \mX {\text{\boldmath$X$}}
\def \yvec {\text{\boldmath$y$}}    \def \mY {\text{\boldmath$Y$}}

 \def \calC {\mathcal C}
 \def \calD {\mathcal D}
 
 \def \calF {\mathcal F}

 \def \calN {\mathcal N}

\def \zhatvec {\text{\boldmath$\hat z$}}

\def \betavec         {\text{\boldmath$\beta$}}

\def \epsilonvec      {\text{\boldmath$\epsilon$}}
\def \varepsilonvec   {\text{\boldmath$\varepsilon$}}

\def \thetavec        {\text{\boldmath$\theta$}}
\def \varthetavec     {\text{\boldmath$\vartheta$}}

\def \mTheta   {\mathbf{\Theta}}

\def \mPhi     {\mathbf{\Phi}}
\def \mPsi     {\mathbf{\Psi}}

\def \nullvec {\mathbf{0}}

\newenvironment{pf}[1][Proof]{\begin{proof}[\textit{\textbf{#1}}]} {\end{proof}} 
\theoremstyle{plain}
\newtheorem{theorem}{Theorem}
 
\newtheorem{proposition}{Proposition} 

\theoremstyle{definition}
\newtheorem{defn}{Definition}

\newtheorem{remark}{Remark}
\title{Phenology curve estimation via a mixed model representation of functional 
principal components: Characterizing time series of satellite-derived vegetation indices}

\author{Inder Tecuapetla-G\'omez$^{(1,2)}$}
\email[Inder~Tecuapetla]{itecuapetla@conabio.gob.mx}
\address{$^1$Programa Investigadoras e Investigadores por M\'exico\\
National Council for Humanities, Science and Technology (CONAHCyT)\\
Mexico City, Mexico}
\address{$^2$Geomatics Unit\\
National Commission for the Knowledge and Use of Biodiversity (CONABIO)\\
Mexico City, Mexico}

\author{Francisco Rosales-Marticorena$^{(3)}$}
\email{frosales@esan.edu.pe}

\address{$^3$Graduate School of Business\\
Universidad ESAN, Lima, Peru}

\author{Berenice Fanny Galicia-G\'omez$^{(4)}$}
\address{$^4$Licenciatura en Actuar\'ia\\
Facultad de Ciencias, UNAM\\
Mexico City, Mexico}
\begin{document}
% ------------------------------------------------------------------------
%
\begin{abstract}
Vegetation phenology consists of studying synchronous stationary events, such
as the vegetation green up and leaves senescence, that can be construed as adaptive
responses to climatic constraints. In this paper, we propose a method to
estimate the annual phenology curve from multi-annual observations of
time series of vegetation indices derived from satellite images.
% we characterize the annual 
%phenology curve based on multi annual vegetation index time series derived from satellite images.
%Although the original data is non-functional, 
We fitted the classical harmonic regression model to annual-based time series in order to 
construe the original data set as realizations of a functional process.
%an application of 
%the classical 
%harmonic regression model allows us to interpret our time series as realizations
%of a functional process. 
Hierarchical clustering was applied to define a nearly 
homogeneous group of annual (smoothed) time series from which a representative and \emph{idealized} 
phenology curve was estimated at the pixel level. 
This curve resulted from fitting a mixed model, based on functional principal components,
to the homogeneous group of time series. 
%novel
%functional PCA-based regression model to the homogeneous group of time series. 
Leveraging the idealized phenology curve, we employed standard calculus criteria
to estimate the following \emph{phenological parameters} (stationary 
events): green up, start of season, maturity, senescence, end of season and
dormancy.
% was carried out allowing us to provide a fair description of the entire
%annual dynamic of remotely sensed vegetation. 
By applying the proposed methodology to four different data cubes (time series from
2000 to 2023 of a popular satellite-derived vegetation index) recorded
across grasslands, forests, and annual rainfed agricultural zones of a Flora
and Fauna Protected Area in northern Mexico, we verified that
%The proposed methodology was applied to four different data cubes (time series from
%2000 to 2023 of a popular satellite-derived vegetation index) recorded
%at the grasslands, forests and annual rainfed agricultural zones of a Flora
%and Fauna Protected Area in northern Mexico; 
our approach characterizes properly the phenological cycle in vegetation
with nearly periodic dynamics, such as grasslands and agricultural areas.
%Our results suggest that the proposed method estimates at least five of the 
%abovementioned phenological parameters successfully. This is in line with some of our 
%simulation studies which suggest that our method may missed some phenological parameters in annual vegetation 
%time series with particularly flattened regions.
%Our methodology is applied to
%three different data cubes (time series of 16-day 250 m vegetation index from 2000 to 2021)
%covering subtropical broadleaf deciduous forests located in some Natural Protected Areas 
%of Mexico. Our results suggest that the
%proposed method estimates at least five of the abovementioned phenological
%parameters successfully. This is in line with some of our simulation studies which 
%suggest that our method may missed some phenological parameters in annual vegetation 
%time series with particularly flattened regions. 
The \texttt{R} package \texttt{sephora} was used for all computations in this
paper.

\end{abstract}
%
% ------------------------------------------------------------------------
%
\maketitle
% 
% the very first time run this:
%\include{Main}
%\include{Supplement}
%\usepackage{chapterbib} and \externaldocument{Supplement} above must be commented

% Once the .aux have been created then run below; dont forget to uncomment
%\usepackage{chapterbib} and \externaldocument{Supplement} and comment \include{Main}
%and \include{Supplement} above, also, uncomment \input{Main} below

% To produce dbe-semiPar-cps_main.pdf

% -----------------------------------------------\textsc{•}-------------------------
\section{Introduction}~\label{sec:intro}
% ------------------------------------------------------------------------

% ---------------------------------------------------------------------------------------
In 1751 the botanist Carolus Linnaeus delineated some techniques for gathering annual 
plant calendars of leaf opening, blooming, fruiting, and leaf fall. Almost a century later,
these concepts were used to coin the term \emph{phenology} by the botanist Charles Morren
in 1853, cf.~\cite{Hopp_1974}.
In the early 1970s the International Biological Program established a Phenology Committee
seeking, among other things, to generate the adequate framework for studying ecosystem's
phenology, \cite{Lieth_1974}. As a result of such efforts, this committee came out with the 
following definition:

\begin{defn}
	Phenology is the study of the timing of recurring biological events, the causes
	of their timing with regard to biotic and abiotic forces, and the interrelation among
	phases of the same or different species.
\end{defn}

Thus, when considering to use this definition in the context of a terrestrial ecosystem,
whose dimension can cover a few to hundreds of hectares,
Earth Observations (EO) from satellite sensors emerge like the best reliable and 
economical source of data for such a purpose. In this work, we focused on estimating
some phenological parameters (recurring biological events occurring at certain time) from
time series of satellite images.

%Thus, with regards to estimating the timing of recurring bioological events on terrestrial
%ecosystems, Earth Observations (EO) from satellite sensors are, arguably, the best reliable and 
%economical source of data for such a purpose. 

In the last three decades we have seen a
sustained interest on applying different mathematical approaches to better understand 
remotely-sensed vegetation dynamics (\cite{reed1994measuring}, \cite{sellers1994global}, \cite{olsson1994fourier},
\cite{roerink2000reconstructing}, \cite{jakubauskas2001harmonic}, \cite{jonsson2002seasonality},
\cite{heumann2007avhrr}, \cite{jonsson2010annual}, \cite{sjostrom2011exploring}, \cite{melaas2013detecting},
\cite{reyes2021cross}). Over these years, a certain level of agreement seems to have been
achieved about the pertinence of using the classical harmonic regression model
to extract the main interannual characteristics of vegetation dynamics when 
this is registered with a coarse resolution data, cf.~\cite{roy2020robust}; 
this data type will be used in this paper applications.

With the use of harmonic regression or other sounding mathematical techniques, it is possible to obtain
a smooth (continuous) version of annual vegetation curves. From these curves
is relatively easy to estimate certain phenological parameters (calendar 
dates at which biological events have occurred) as the critical points of some
of their derivatives (\cite{zhang2003monitoring}, \cite{klosterman2014}), and subsequent 
studies of the interannual dynamics of these parameters can be obtained (\cite{boschetti2009multi}, 
\cite{Eklundh_2015}, \cite{zheng2016crop}, \cite{zeng2020review}).

Due to the release of time series of EO with mid and high resolution,
such as Landsat and Sentinel, some researchers have proposed to combine -the fusion of coarse and mid,
coarse and high or mid and high resolution images- these datasets and apply different analytical 
methods to produce annual vegetation phenology curves (\cite{zeng2016hybrid}, 
\cite{filippa2016phenopix}, \cite{baumann2017}, \cite{liao2019using}, \cite{zhao2021}, \cite{nietupski2021}, \cite{sisheber2022}).
In this paper, we propose an approach to estimate the annual vegetation curve intrinsic
to the long-term dynamics of (pixel-based) time series of NASA's Moderate Resolution 
Imaging Spectroradiometer (MODIS) images. 

%MODIS images have been available, free of cost, since February 18, 2000. In particular, the 
%MODIS product MOD13Q1 provides pre-processed images of the Normalized Difference Vegetation 
%Index (NDVI); the version 6.1 of this product was released in July 2023.
%Due to its temporal resolution (16 days) 
%a typical annual NDVI MOD13Q1 time series has 23 observations, and consequently, a full time 
%series (2000-2023) of this product has 549 observations, an amount that is inherent to any 
%pixel of this time series. Thus, at the pixel level, we applied the classical harmonic
%regression model to the observations of each of the 24 annual NDVI time series, producing 
%continuous and periodic functions. Then, we applied a typical functional principal 
%component analysis (FPCA) to this set of functions and
%used a finite number of these FPCs to represent an idealized NDVI curve. HERE we have to write something about FPCA and conclude with \emph{to the best of our knowledge
%FPCA-based applications to recover the phenology curve have not yet been explored}.

MODIS images have been available, free of cost, since February 18, 2000. In particular, the 
MODIS product MOD13Q1 provides pre-processed images of the Normalized Difference Vegetation 
Index (NDVI); the lattest version (6.1) of this product was released in July 2023.
Due to its temporal resolution (16 days), a full-length NDVI MOD13Q1 time series (2000-2023)
has 549 observations. At the pixel level, we applied the classical harmonic regression 
model to the original observations and map them into 24 continuous and periodic functions. Then, we applied a typical functional principal component analysis (FPCA) to this 
set of functions and used a finite number of these FPCs to represent an idealized 
NDVI curve; the estimation of this idealized curve is performed via a mixed model.

%HERE we have to include some relevant and actual references about FPCA and conclude 
%with \emph{to the best of our knowledge FPCA-based applications to recover the 
%phenology curve have not yet been explored}.

Functional data analysis (FDA) has emerged as a crucial tool in various real-life 
applications due to its ability to handle complex, high-dimensional data structures 
commonly encountered in fields such as biology, economics, engineering, and environmental science (\cite{Ramsay_Silverman_2005}; \cite{ferraty2006}; \cite{kokoszka2017}). 
By treating data as functions rather than discrete points, FDA allows researchers to capture 
and analyze intricate patterns and relationships over continuous domains (\cite{Ramsay_Silverman_2005}). This approach enables a deeper understanding of dynamic 
systems and processes, 
leading to more informed decision-making and improved predictive modeling. For instance, 
in ecology and environmental science, FDA has been instrumental in analyzing spatiotemporal trends in vegetation dynamics (\cite{ferraty2006}; \cite{kokoszka2017}; \cite{Yu2020}), while in economics, it has facilitated the modeling of time-varying risk factors in financial markets (\cite{bai2002determining}; \cite{Ramsay_Silverman_2005}). One powerful technique within FDA is functional principal component analysis (FPCA), which extracts dominant modes of variation 
from functional data, allowing for dimensionality reduction and efficient representation 
of variability (\cite{Ramsay_Silverman_2005}; \cite{kokoszka2017}). Applications of FPCA 
range from characterizing climate patterns (\cite{lopez2009concept}) to analyzing brain 
imaging data (\cite{tian2010functional}), showcasing its versatility and effectiveness in diverse domains. To the best of our knowledge FPCA-based applications to recover the 
phenology curve from satellite images have not yet been explored.

%a typical annual NDVI MOD13Q1 time series has 23 observations, and consequently, a full time 
%series (2000-2023) of this product has 549 observations, an amount that is inherent to any 
%pixel of this time series. Thus, at the pixel level, we applied the classical harmonic
%regression model to the observations of each of the 24 annual NDVI time series, producing 
%continuous and periodic functions. Then, we applied a typical functional principal 
%component analysis (FPCA) to this set of functions and
%used a finite number of these FPCs to represent an idealized NDVI curve. HERE we have to write something about FPCA and conclude with \emph{to the best of our knowledge
%FPCA-based applications to recover the phenology curve have not yet been explored}.

We construe this idealized NDVI curve as a fair approximation to the vegetation phenology curve
intrinsic to a pixel. From this curve, we utilized basic calculus' criterion to find
phenological parameters such as Green up (GU), Start of Season (SoS), Maturity (Mat), 
Senescence (Sen), End of Season (EoS) and Dormancy (Dor) which provide an appropriate characterization
of the (remotely-sensed) vegetation cycle. %\cite{Tecuapetla_etal_2023}

This paper is organized as follows. In Section~\ref{sec:methods} we introduced our
methodology as well as a mathematically-oriented definition of the phenological parameters mentioned
above, see Table~\ref{tab:phenopars}. Based on this definition and certain conditions, we established that
SoS and EoS are always achieved, see Theorem~\ref{theo:start_end_season}. 
Also in this section, we considered the simplest phenology 
curve (based on harmonics) and found GU, SoS, Mat and EoS explicitly,
see Proposition~\ref{prop:singleHarmonics}.
Section~\ref{sec:sims} presents three simulation studies assessing the precision of our methods
for estimating phenological dates. In Section~\ref{sec:apps}, we considered
the MOD13Q1 v6.1 NDVI time series (2000-2023) corresponding to the Flora and Fauna Protected Area \emph{Cerro Mohinora} located at Chihuahua, Mexico; we applied our methodology to describe the phenological cycle
in four different land covers: Grasslands, Douglas Fir forest, Pine-Oak forest and Annual Rainfed
Agriculture.

%Although these sounding technical principles have been employed to estimate vegetation annual curves,
%the estimation of certain phenological parameters (calendar 
%dates at which biological events have occurred) have been carried out with less unambiguous
%techniques (\cite{Eklundh_2015}). More recently, 

% ------------------------------------------------------------------------
%
% ------------------------------------------------------------------------
\section{Methods and Main Results}~\label{sec:methods}
% ------------------------------------------------------------------------
In this section we present a procedure to compute phenological parameters
from long-term NDVI time series (pixel-based) such as the MOD13Q1 product.
%in the NDVI process. We use the natural protected area data described in section 2, 
%which contains NDVI information at the pixel level between the years 2000 and 2021. 
%Our procedure is purely pixel-based.
%For each pixel, 
Conceptually, we begin by disjoining the time series in annual segments that can be 
seen as functional samples of an underlying NDVI process; this idea is similar in spirit
to a profile analysis. Operatively speaking, we begin by representing the 
annual time series data as a function by fitting a harmonic regression. 
%We continue by properly disjoining the resulting function in yearly segments that can be 
%seen as functional samples of an underlying NDVI process; this idea is similar in spirit
%to a profile analysis. 
We then apply a statistical algorithm -based on FPCA-
%, based on functional principal component analysis (FPCA), 
to extract an annual functional signature from the functional NDVI samples, 
independent of anomalous annual specific variations.
Finally, basic calculus criteria are used to identify phenological dates from the
resulting signature.

%; this is achieved by employing a statistical algorithm to the resulting signature 
%to identify the phenology curve inherent to a pixel. 

%=================================================================================
\subsection{Functional data} 
%=================================================================================

Functional data refers to data observed as functions over a continuous domain. Unlike traditional data, which are typically represented as a set of discrete measurements, functional data are represented as curves or smooth functions, see e.g., \cite{ramsay} \cite{ferraty}, \cite{kokoszka} among many others. This representation allows for the exploration of the inherent structure and variability within the data, capturing the underlying structure of the functional process. To this end, for each pixel  in the data set, we first fit the time series data to a harmonic function. We then build a set of functional samples, and remove outlier functional observations.
  
%----------------------------------------------------------------------------------------------------------------------------------------------
\subsubsection{Harmonic functions} 
%----------------------------------------------------------------------------------------------------------------------------------------------

Consider the sequence $\{z_t\}_{t=1}^L$ of NDVI measurements at a given location, taken at a given time $t$ within certain year, and a characterization of the form 
\begin{align}%\label{eq:harmodel}
    z_{t} 
    &= 
    g(t) + \varepsilon_t,\quad \varepsilon_t 
    = \sigma^2\,Z,  
    \quad Z\sim N(0,1),\label{eq-repeatedTimeSeries} \\ 
    g(t) 
    &= 
    \theta_0 
    + 
    \sum_{j=1}^p\,
    \left\{ \alpha_j\,\sin\left( \frac{2\pi\,j\,t}{L} \right)
    + 
    \beta_j\,\cos\left( \frac{2\pi\,j\,t}{L} \right) \right\},
    \quad t=1,2,\ldots,L,\label{eq-harmRegression}
\end{align}
where  $g$ is assumed to be a continuous periodic function, and thus suitable to be approximated by the addition of various sine and cosine functions of varying amplitude ($\alpha_j$, $\beta_j$),  and period ($2\pi j/L$).  To the best of our knowledge most of the empirical studies on NDVI  time series have considered 
% $\sigma_t^2=\sigma^2$ for all $t$, and suggested to use 
$p=3$ or $p=4$, cf.~\cite{eastman}.

Model \eqref{eq-repeatedTimeSeries} can be written in matrix form as
\begin{equation}\label{eq:matrixform}
	\bs{z} 
	= 
	\mX\betavec 
	+ 
	\varepsilonvec, \quad\varepsilonvec\sim N(\nullvec,\sigma^2\mI),
\end{equation}
where $\bs{z},\varepsilonvec\in\mathbb{R}^L$, $\betavec\in\mathbb{R}^{2p+1}$, 
and  $\mX$ is the design  matrix $(L\times (2p+1))$ of the form
\begin{equation}
    	\mX=
	\left[
	\begin{matrix}
    	1 & \sin\left( \frac{2\pi}{L}  \right) & \cos\left( \frac{2\pi}{L}  \right)& \dots & \sin\left( \frac{2p\pi}{L}  \right)& \cos\left( \frac{2p\pi}{L}  \right)\\
    	1 & \sin\left( \frac{4\pi}{L}  \right) & \cos\left( \frac{4\pi}{L}  \right)& \dots & \sin\left( \frac{4p\pi}{L}  \right)& \cos\left( \frac{4p\pi}{L}  \right)\\
	\vdots&\vdots&\dots&\vdots&\vdots\\
	1 & \sin\left( {2\pi}  \right)& \cos\left( {2\pi}  \right) & \dots & \sin\left( {2p\pi}  \right)& \cos\left( {2p\pi}  \right)	
    	\end{matrix}
	\right].
\end{equation}
Estimation of \eqref{eq:matrixform} requires identification of parameter vector $\varthetavec=\{\theta_0,\alpha_1,\dots,\alpha_p,\beta_1,\dots,\beta_p,\sigma\}$, 
and  selection of hiper-parameter $p$. Model~\eqref{eq:matrixform} is applied
to each set of annual observations; for our application, we ended up with 24 smoothed NDVI annual functions. 

%These goals can be achieved by numerical minimization of the AIC criteria, defined as 
%	\begin{equation}
%	\mbox{AIC} 
%	&=& 2(p+1) - 2\,\log(\calL({\varthetahatvec}; p)),\\
%    	\calL({{\varthetahatvec}};p) &\propto& \prod_{i=1}^L\,\sigma_i^{-1}\, 
%    	\mbox{exp} \left\{ -\frac{1}{2}\, \sum_{i=1}^L\, \left(\frac{ z_i - \hat z_i}{\sigma_i}\right)^2 \right\},
%	\end{equation}
%where $\calL({{\varthetahatvec}};p) $ is the likelihood maximized at the parameter $\hat{\varthetavec}$, given $p$, and  $\zhatvec=\mX\betahatvec$.

%----------------------------------------------------------------------------------------------------------------------------------------------
%\subsubsection{Preprocessing Functional Data} 
\subsubsection{Hierarchical clustering}
%----------------------------------------------------------------------------------------------------------------------------------------------

We apply hierarchical clustering, cf.~\cite{hcluster}, to the set of NDVI smoothed functions
obtained in the previous section. We utilize the function \texttt{tsclust} of the R package
\texttt{dtwclust} by \cite{dtwclust} as the main engine to perform clustering. We set 
\texttt{tsclust} to find two clusters only. By doing so, we aim to find a subset of functions 
which are similar -close in distance- and at the same time discard from the functional dataset
those curves with an atypical behavior. Although this step intends to facilitate
the application of the upcoming FPCA algorithm, by pre-selecting a set of highly similar curves,
our implementation also allows for the use of either of the two clusters or the entire
functional dataset in the FPCA step.

%included and outlier curves are removed.
%To clean the data and  facilitate the application of the FPCA algorithm at the pixel level, we follow the construction of the NDVI fit with a yearly segmentation of it, thus building a set of functional samples. We then apply  hierarchical clustering \cite{hcluster} per pixel, such that only a subset of similar representative curves is included and outlier curves are removed. 

%We highlight that we do not align our data set, e.g. performing time warping or similar methods, as our goal is to identify each pixel's  phenological dates, and these are characterized in physical time. 

%=================================================================================
\subsection{Functional principal components}
%=================================================================================
A random variable $\calF$ is called a functional variable if it takes 
values in an infinite dimensional space (or functional space). Since $\calF$ is a random curve, we can identify it as $\calF=\{\calF(t); t\in[0,1] \}$. 
Functional data  analysis (FDA) takes place in a functional data set   $f_1,\dots,f_m$ of  realizations of $m$  functional variables $\calF_1,\dots,\calF_m$, identically distributed as $\calF$, where an observation $f$ of $\calF$ is called ``functional data".  We highlight that we assume  the functional observations $f_1,\dots,f_m$ are independent. 
The interested reader is refered to \cite{kokoszka} to explore the dependence 
case. 

Our functional data is given as a collection of functions sampled at $n$ equidistant points and arranged as a rectangular matrix with $m$ columns and $n$ rows. Let $\mY=(\zhatvec_1^\top,\zhatvec_2^\top,\dots,\zhatvec_m^\top)^\top$, 
where each functional observation is captured in the columns of $\mY$ and corresponds to the estimation of different samples of NDVI  at the same location for different years using model \eqref{eq-repeatedTimeSeries}-\eqref{eq:matrixform}. 
The goal of FPCA is to identify the  functional principal components that explain most of the variance in matrix $\mY$, and use them to build an overall trend curve $\tau$, that represents the NDVI signature at a given location, regardless of the year.  In this subsection we present the general theory of FPCA and the 
corresponding estimation algorithm. 

%----------------------------------------------------------------------------------------------------------------------------------------------
\subsubsection{Principal components} \label{subsec:fpca}
%----------------------------------------------------------------------------------------------------------------------------------------------

Consider   function $f_j(t)$ for $j=1,2,\dots,m$, defined 
in the compact interval $[0,1]$  as functional data. More specifically, assume each $ f_j(t)$ is an independent realisation 
of the stochastic process $\{f(t), t\in[0,1]\}$ with mean $\mathbb{E}[ f(t)]=\tau(t)$ and covariance kernel 
$\mathcal{K}(x,z)=\mbox{cov}\{ f(t), f(z)\}$, $x,z\in [0,1]$. Mercer's lemma states that if 
$\int_0^1\mathcal{K}(x,x)dx<\infty$ then there exists an orthonormal sequence of eigenfunctions $\{\zeta_j\}_{j=1}^\infty$ and 
non-increasing, non-negative sequence of eigenvalues $\{\kappa_j\}_{j=1}^\infty$ such that for 
$
(\mathcal{K}\zeta_j)(t):=\int_0^1\mathcal{K}(x,z)\zeta_j(z)dz=\kappa_j\zeta_j(t),
$
it holds that 
\[
\mathcal{K}(x,z)
=
\sum_{j=1}^\infty\kappa_j\zeta_j(t)\zeta_j(z),
\mbox{ and }\sum_{j=1}^\infty\kappa_j=\int_0^1\mathcal{K}(x,x)dx.
\]
%~\\

Hence we can write the Karhunen-Lo\`eve expansion as 
\begin{equation}\label{eq:kl}
 f(t)=\tau(t)+\sum_{j=1}^\infty\sqrt{\kappa_j}\xi_j\zeta_j(t),
\end{equation}
where $\xi_j:=\frac{1}{\sqrt{\kappa_j}}\int{f}(t)\zeta_j(z)dz$, $\mathbb{E}[\xi_j]=0$, $\mathbb{E}[\xi_j,\xi_k]=\delta_{jk}$, $j,k\in\mathbb{N}$ and $\delta_{j,k}$ is the Kronecker delta. From (\ref{eq:kl}) the implementation of a reduced rank model for a sample $j=1,2,\dots,m$ as presented in e.g. \cite{james} can be written as 
\begin{equation}\label{eq:rr}
f_j(t)=\tau(t)+\sum_{k=1}^h \psi_k(t)v_{jk}=\tau(t)+\mPsi(t)^\top\vvec_j
\end{equation}
where $h$ is a finite integer usually $h\ll m$, $\mPsi(t)=\{\psi_1(t),\psi_2(t),\dots,\psi_h(t)\}^\top$ contain the $k$-th principal component (PC) function $\psi_k(t)$ and $\vvec_j=\{v_{j1},v_{j2},\dots,v_{jh}\}^\top$ is the vector of PC scores for the $j$-th curve. Moreover, we can interpret each $f_j(t)$  in (\ref{eq:rr}) as a curve composed by an overall trend $\tau(t)$  and a subject specific deviation $\eta_j(t):=\sum_{k=1}^h \psi_k(t)v_{jk}$.

Thus, given observation pairs $(t_{i},y_{i,j})$ for $i=1,\dots,L$ and $j=1,\dots,m$, one can write an additive model of the form 
\begin{equation}\label{eq:signals}
y_{i,j}=\tau(t_i)+\eta_j(t_i)+\epsilon_{i,j},
\end{equation}
where $\tau(t)$ is an unknown smooth function and  $\{\epsilon_{i,j}\}_{i=1}^n$ is the uncorrelated homoscedastic error measurement of curve $j$, i.e. $\Cor(\epsilon_{i,j},\epsilon_{k,j})=\ind_{\{i=k\}}$ for any $(i,j)$ pair. Given a unique smoothness class $q=2$ for all curves, see \cite{Krivobokova_etal_2022} for details, we can represent the unknown functions as 
$\tau(t)=\bs{C}_\tau(t)\bs{\theta}_\tau$ and $\psi_j(t)=\bs{C}_\psi(t)\bs{\theta}_{\psi_j}$. For the last case we can also write $\mPsi(t)^\top=\bs{C}_\psi(t)\bs{\Theta}_\psi$ with $\bs{\Theta}_\psi=\{\thetavec_{\psi_1},\dots,\thetavec_{\psi_h}\}$ an $n\times h$ matrix such that
\[
\int \mPsi(t)\mPsi(t)^\top dx
=
\bs{\Theta}_\psi^\top\left(\int \bs{C}(t)^\top\bs{C}(t) dx\right)\bs{\Theta}_\psi
=
\mI_h,
\]
holds, and the usual orthogonality requirements for the principal component curves are satisfied. It follows that  $\bs{\Theta}_\psi^\top\bs{\Theta}_\psi=\mI_h$  and thus the estimation problem in (\ref{eq:rr}) is reduced to the computation of the spline 
coefficients $\thetavec_\tau$,  $\thetavec_{\psi_k}$ for $j=1,\dots,m$, and $k=1,\dots,h$ in 

\begin{eqnarray}\label{eq:minproblem}
\min_{\thetavec_\tau, \thetavec_{\psi_k}}&&
\left[\frac{1}{n}\sum_{j=1}^m(\yvec_j-\mC_\tau\thetavec_\tau-\mC_{\psi}\mTheta v_j)^\top(\yvec_j-\mC_\tau\thetavec_\tau-\mC_{\psi}\mTheta \vvec_j)\right.\nonumber\\ 
&&+\left.\lambda_\tau\thetavec_\tau^\top
\left(
\int_0^1\left\{\mC_\tau(t)^{(2)}\right\}^\top\left\{\mC_\tau(t)^{(2)}\right\}dt
\right)\thetavec_\tau\right.\nonumber\\
&&+\left.\lambda_{\psi}\sum_{k=1}^{h}\thetavec_{\psi_k}^\top
\left(
\int_0^1\left\{\mC_{\psi}(t)^{(2)}\right\}^\top\left\{\mC_{\psi}(t)^{(2)}\right\}dt
\right)\thetavec_{\psi_k}\right], 
\end{eqnarray}
where $\vvec_j$ is a random vector such that  $\vvec_j\sim{\calN}\left(\bm{0},\sigma_{v_j}\mI_{m}\right)$, we denote $\mPsi=\mC_\psi\mTheta_\psi$ and  $\mTheta^\top\mTheta=\mI_h$. Given $h$ it remains to estimate  $\vvec_j$, $\lambda_\tau$ and $\lambda_\psi$ for which one can utilize the mixed models representation of penalized splines, see e.g. \cite{rupert}. 
The estimation of the full additive model as in \eqref{eq:signals} it is implemented as part
of the function \texttt{phenopar} of our R package \texttt{sephora}.

%----------------------------------------------------------------------------------------------------------------------------------------------
\subsubsection{Statistical algorithm} \label{subsec:fpca.algo}
%----------------------------------------------------------------------------------------------------------------------------------------------
Here we present the two main steps of the statistical algorithm implemented to 
identify $\tau(t)$ in \eqref{eq:signals}. 

\underline{Step 1:}\\
\begin{enumerate}
\item Solve (\ref{eq:minproblem}) for a trend only model, i.e. for $\bs{C}_{\psi}\bs{\Theta} v_j=0$ for all $j$, to
obtain $\lambda_\tau$ and compute 
\begin{equation}
\hat\thetavec_\tau^{(0)}
=
\left(\calC_\tau^\top\calC_\tau+\lambda_\tau n\calD_\tau\right)^{-1}
\calC_\tau^\top\mY
\end{equation}
with $\calC_\tau=\bs{C}_\tau\otimes\mI_m$ and $\calD_\tau=\bs{D}_\tau\otimes\mI_m$.
\item Write the residuals for each curve as $\bs{r}_j=\bs{y}_j-\bs{C}_\tau\hat\thetavec_\tau^{(0)}$ and fit the linear least squares model
\[
\bs{r}_j=\bs{C}_\psi\bs{\Gamma}_j+\epsilonvec_j,
\]
to obtain 
$\hat{\bs{\Gamma}}^{(0)}=(\hat{\bs{\Gamma}}_1^{(0)},\ldots,\hat{\bs{\Gamma}}_m^{(0)})^\top$.
%$\hat{\bs{\Gamma}}^{(0)}=\{\hat{\bs{\Gamma}}_1^{(0)},\dots,\hat{\bs{\Gamma}}_m^{(0)}\}^\top$.
\item Calculate the singular value decomposition 
\[
\hat{\bs{\Gamma}}^{(0)}=\bs{U}\bs{\Sigma}\bs{V}^\top,
\]
and set $\hat\thetavec_{\psi_k}^{(0)}=\bs{V}_k\bs{\Sigma}_k$, and 
hence $\hat{\mPsi}^{(0)}=\bs{C}_\psi\hat{\bs{\Theta}}_\psi^{(0)}$.
\end{enumerate}
~\\
\underline{Step 2:}\\ 
~\\
Given $h$, initial value $\hat{\mPsi}^{(0)}$ and parametrisation $\mPhi$, 
update the computation of $\hat{\bs{v}}^{(l)}$ in (\ref{eq:minproblem}) for the $l$-th iteration and the corresponding $\hat{\mPsi}^{(l)}$ matrix until convergence is achieved. 

\begin{enumerate}
\item Solve (\ref{eq:minproblem}) given $\hat\mTheta_\psi^{(l-1)}$ and update $\hat\vvec^{(l)}$ and $\hat\lambda_\tau^{(l)}$ to compute
\begin{equation*}
\hat{\thetavec}_\tau^{(l)}
=
\left\{\calC_\tau^\top\calC_\tau+\hat\lambda_\tau^{(l)} n\calD_\tau\right\}^{-1}\calC_\tau^\top\mY,%\nonumber
\end{equation*}
with $\calC_\tau$ and $\calD_\tau$ as defined on the initialisation step.
\item Write the residuals  as $\rvec^{(l)}=\mY-\calC_\tau\hat\thetavec_\tau^{(l)}$ and fit the penalised least squares model
$$
\rvec^{(l)}=\mC_\psi\hat\mTheta_\psi^{(l-1)}\hat\vvec^{(l)}+\epsilonvec,
$$
to update $\hat\mTheta_\psi^{(l)}$ by
\begin{equation}
\hat{\thetavec}_{\psi_k}^{(l)}=
\left\{
\sum_{j=1}^N \left(\hat{v}_{jk}^{(l)}\right)^2\mC_\psi^\top\mC_\psi+\hat{\lambda}_\psi^{(l)}\mD_\psi
\right\}^{-1}%\nonumber\\
\left\{
\sum_{j=1}^N \left(\hat{v}_{jk}^{(l)}\right)\mC_\psi^\top\left(\mY_j-\mC_\tau\hat{\thetavec}_\tau^{(l)}-\mC_\psi\hat\mQ_{jk}^{(l)}\right)
\right\}\nonumber,
\end{equation}
for 
$$
\hat{\mQ}_{jk}^{(l)}=\sum_{l\neq k}\hat\thetavec_{\psi_l}^{(l)}\hat v_{jl}^{(l)},\;\; j=1,\dots,N.
$$ 
\item Construct $\hat{\mTheta}_\psi^{(l)}$ and use the QR decomposition to orthonormalise its columns. 
With the new estimation of $\hat{\mTheta}_\psi^{(l)}$ go back to step one until convergence is achieved.

\end{enumerate} 

%=================================================================================
\subsection{Phenological date estimation}
%=================================================================================

We propose a formal definition of phenological date estimates which is
based on the second derivative criterion from basic calculus.
Let $\tau:\calD\to\mathbb{R}$ denote the idealized NDVI signature obtained via FPCA. 
Table \ref{tab:phenopars} shows our definition in an schematic manner.
Our definition, involves the first four derivatives of $\tau(x)$, 
$x\in\calD$, and we believe that complements the treatment of phenological dates presented 
in \cite{baumann2017}.

%The proposed method identifies phenologicadates, as constructs of critical points 
%of certain derivatives of $\tau$, which are readily available from 
%the solution of \ref{eq:minproblem}. 
%Point-wise confidence bands for $\tau$ and its derivatives are also provided. 

%----------------------------------------------------------------------------------------
%\subsubsection{Critical Point Based Criteria}
%----------------------------------------------------------------------------------------

%In Table \ref{tab:phenopars} we propose a definition of phenological dates using basic calculus
%second derivative criterion. Our definition, involves the first four derivatives of $\tau(x)$, 
%$x\in\calD$, and we believe that complements the treatment of phenological dates presented in \cite{baumann}.

%\begin{table}[htbp]
%\caption{Characterization of Phenological Dates}
%\begin{center}
%\label{tab:fs}
%\begin{tabular}{l|l|l|cccc}
%\hline 
%Date 				&Name			&Critical Point		&$\tau^{(1)}$		&$\tau^{(2)}$		&$\tau^{(3)}$			&$\tau^{(4)}$\\
%\hline
%$x_{\mbox{\tiny GUp}}$	&Green-up 		&$\tau^{(2)}$ maxima	&+			&+			&0				&-\\
%$x_{\mbox{\tiny SoS}}$	&Start of Season	&$\tau^{(1)}$ maxima	&+			&0			&-				&\\
%$x_{\mbox{\tiny Mat}}$	&Maturity			&$\tau^{(2)}$ minima	&+			&-			&0				&+\\
%\hline
%$x_{\mbox{\tiny Sen}}$	&Senescence		&$\tau^{(2)}$ minima	&-			&-			&0				&+\\
%$x_{\mbox{\tiny EoS}}$	&End of Season	&$\tau^{(1)}$ minima	&-			&0			&+				&\\
%$x_{\mbox{\tiny Dor}}$	&Dormancy		&$\tau^{(2)}$ maxima	&-			&+			&0				&-\\
%\hline
%\end{tabular}
% \end{center}
% \end{table}

\begin{table}[h]
\caption{Definition of phenological parameters}\label{tab:phenopars}
\centering
\scalebox{0.85}{
\begin{tabular}{cccccc}
\toprule[1.25pt]
            Phenological parameter & Notation & Critical point & $\tau^{(2)}$ & $\tau^{(3)}$ & $\tau^{(4)}$ \\
\cmidrule[1.25pt]{4-6}
          Green up & $x_{GU}$  & $\tau^{(2)}$ (global) maximum & $\cdot$ & 0 & - \\
   Start of season & $x_{SoS}$ & $\tau^{(1)}$ (global) maximum & 0 & - & $\cdot$ \\
          Maturity & $x_{Mat}$ & $\tau^{(2)}$ (global) minimum & $\cdot$ & 0 & + \\
        Senescence & $x_{Sen}$ & $\tau^{(2)}$ (local) minimum & $\cdot$ & 0 & + \\
     End of season & $x_{EoS}$ & $\tau^{(1)}$ (global) minimum & 0 & + & $\cdot$ \\
          Dormancy & $x_{Dor}$ & $\tau^{(2)}$ (local) maximum & $\cdot$ & 0 & - \\
\bottomrule[1.25pt]  
\end{tabular}
}
\end{table}

In addition to Table \ref{tab:phenopars}'s definition, the general -and natural- constraint must be imposed:

%Note that in table \ref{tab:fs} each phenological date is associated to a critical point 
%in one of the derivatives of $\tau(x)$, but to uniquely identify each date additional conditions are required. For example, Green-up ($x_{\mbox{\tiny GUp}}$) 
%and  Dormancy ($x_{\mbox{\tiny Dor}}$) are both local maxima of $\tau^{(2)}$ where $\tau^{(2)}$ is positive, and are only different from one another in that: i) 
%for $\tau^{(1)}(x_{\mbox{\tiny GUp}})>0$ while $\tau^{(1)}(x_{\mbox{\tiny Dor}})<0$; and that   $x_{\mbox{\tiny GUp}}$ comes before   $x_{\mbox{\tiny SoS}}$, while 
%$x_{\mbox{\tiny Dor}}$ comes after it. That is, the general ordering 
\begin{equation}
	x_{\mbox{\tiny GU}} <
	x_{\mbox{\tiny SoS}}<
	x_{\mbox{\tiny Mat}}<
	x_{\mbox{\tiny Sen}}<
	x_{\mbox{\tiny EoS}}<
	x_{\mbox{\tiny Sen}}<
	x_{\mbox{\tiny Dor}}.
\end{equation}

Thus, it is immediate that

\begin{theorem}\label{theo:start_end_season}
	Suppose that $\tau$ is the mean trend function of the additive model
	\eqref{eq:signals}. Then, the phenological
	parameters Start and End of season always exist on the interval $[0,L]$.
	Also, either Green up or Dormancy and Maturity or Senescence always exist
	on the interval $[0,L]$.
\end{theorem}
\begin{pf}
	The existence of Start and End of Season is ensured by the continuity of
	$\tau^{(1)}$ on $[0,L]$. Similary, the continuity of $\tau^{(2)}$ on $[0,L]$
	ensures that either Green up or Dormancy (and Maturity or Senescence)
	exist.
\end{pf}

One limitation of our method is that the precision of the estimation of $\tau^{(\nu)}$ 
decreases in $\nu$, thus the estimation of Start and End of season, which only require 
the estimation of the second and third derivative of $\tau$, is expected to be more 
precise than that of any other phenological date. 

In order to provide a better understanding of the limitations of our method,
we consider the case of estimating the phenological parameters on the
simplest of harmonic functions: the cosine function.

\begin{proposition}\label{prop:singleHarmonics}
	Let $\tau$ be the mean trend function of the additive model
	\eqref{eq:signals}. Suppose further that $\tau$ follows the representation
\begin{equation}\label{eq:basicSignal}
	\tau(t) = c_0 + c_1\,\cos\left( \frac{2\pi t}{L} - \varphi_1 \right),
\end{equation}
where $c_0, c_1\in \R^{+}$ and $\varphi_1\in (0, 360)$. Then
\begin{align*}
  x_{GU}
  &=
  \frac{(\varphi_1-180)L}{360}
%  (\varphi_1-180)\left( \frac{L}{360} \right)
  \,\ind_{[180,\infty)}(\varphi_1)\\
  x_{SoS}
  &=
  \frac{(\varphi_1-90)L}{360}
%  (\varphi_1-90)\left(\frac{L}{360}\right)
  \,\ind_{[90,\infty)}(\varphi_1)\\
  x_{Mat}
  &=
  x_{Sen}
  =
  \frac{L}{360}\,\varphi_1\\
  x_{EoS}
  &=
  \frac{(\varphi_1+90)L}{360}
%  (\varphi_1+90)\left(\frac{L}{360}\right)
  \,\ind_{[0,270)}(\varphi_1)\\
  x_{Dorm}
  &=
  \frac{(\varphi_1+180)}{360}
%  (\varphi_1+180)\left( \frac{L}{360} \right)
  \,\ind_{[0,180)}(\varphi_1).
\end{align*}
\end{proposition}

\begin{remark}
We have set $x_{Sen}$ equal to $x_{Mat}$ because
$\tau^{(2)}$ has only one minimum on $[0,L]$. In order for
$x_{SoS}$ and $x_{EoS}$ to be found on the interval $[0,L]$,
it is necessary that $90\leq \varphi_1 < 270.$ Observe also
that if $\varphi_1\in [180,270)$, then $x_{GU}$ is well-defined
on the interval $[0,L]$ but, in this case, $x_{Dor}$ cannot
be found on the same interval. Similarly, if $\varphi_1 \in [90,180)$, then
$x_{Dor}$ is found on the interval $[0,L]$ at the expense
of not being able to find $x_{GU}$.
\end{remark}

According to this proposition, the estimates of phenological
parameters are in correspondence with the phase angle $\varphi_1$.
Also, from the remark it follows that with a single harmonics,
is possible to estimate 4 phenological parameters at most:
$x_{GU}$, $x_{SoS}$, $x_{Mat}=x_{Sen}$ and $x_{EoS}$ if $\varphi \in \varphi_1\in [180,270)$;
and $x_{SoS}$, $x_{Mat}=x_{Sen}$, $x_{EoS}$ and $x_{Dor}$ if $\varphi \in \varphi_1\in [90,180)$.

\section{Simulations}~\label{sec:sims}
%% ------------------------------------------------------------------------

% -----------------------------------------------------------------------------

In this section we present a set of simulation studies aiming to assess the precision
of our methodology for estimating the so-called phenological parameters as
they are defined in Table~\ref{tab:phenopars}. 

\subsection{sephora: Well-implemented routines}\label{ssec:sims-intro}

We begin this section by providing evidence that the routines that will be
employed in the upcoming simulation studies have been implemented correctly. To this end,
we consider the signal described in Eq.~\eqref{eq:basicSignal} with $c_0=0$,
$c_1=1$, $L=23$, $\varphi=210$ and $t\in (0, 23)$. We produced 23 observations
from this curve. Then, we fitted the corresponding harmonic regression model (via the
function \texttt{haRmonics} of the R package \texttt{geoTS} by \cite{geoTS})
to these observations; this allows us to get a numerical function from which (and its derivatives)
we can get critical points (phenological parameters). Next, we applied the functions 
\texttt{ndvi\_derivatives} (to get the first fourth derivatives of the numerical function just mentioned), 
\texttt{global\_min\_max} (to get green-up, start of season, maturity and end of season) 
and \texttt{local\_min\_max} (to get senescence) of the R package \texttt{sephora}.
Finally, we obtained the results showed in the Table~\ref{tab:sephora-estimates}.

\begin{table}[htb]
\caption{\label{tab:sephora-estimates} Comparison of theoretical phenological parameters as
provided by Proposition~\ref{prop:singleHarmonics} and the estimates given by the
R package \texttt{sephora}.}
\centering
\scalebox{0.65}{
\begin{tabular}{c|rr}
\toprule
Phenological parameters & Theoretical solutions & \texttt{sephora} estimates \\
\midrule
 GU & 1.916667 & 1.916667\\
SoS & 7.666667 & 7.666667\\
Mat & 13.416667 & 13.416667\\
EoS & 19.166667 & 19.166667\\
\bottomrule
\end{tabular}}
\end{table}

In what follows, we will show results based on the use of \texttt{sephora}'s \texttt{phenopar} function.
This function depends internally on \texttt{haRmonics}, \texttt{ndvi\_derivatives},
\texttt{global\_min\_max} and \texttt{local\_min\_max}, among many other functions. 
In addition to the beginning and end of the time series, \texttt{phenopar} requires
the arguments: \texttt{numFreq} (the number of frequencies employed by \texttt{haRmonics} in the 
smoothing step), \texttt{distance} (the functional distance
employed in the hierarchical clustering step), \texttt{k} (the number of functional principal components),
\texttt{basis} and \texttt{samples} (these 2 arguments are used to calculate the Demmler-Reinsch basis
numerically, the length of each member of the basis is equal to \texttt{samples}).

Unless specified otherwise, we used \texttt{numFreq=1}, \texttt{k=1}, \texttt{samples=50}
and since we are interested in measuring the robustness of our methodology
to the distance used in the hierarchical clustering step, for each simulated time series,
the phenological parameters were estimated under \texttt{distance=dtw\_basic} and \texttt{distance=dtw2}.

\subsection{Homoscedastic error noise}

 In order to assess the performance of our methodology on time series whose length is 
comparable with that of our applications' time series, we replicated the base signal described 
in the previous section (with the same parameters) 23 times; we ended up with a signal 
with 552 observations. To this signal we added randomly generated numbers distributed Gaussian with
zero mean and standard deviation $\sigma \in \{0.125, 0.25, 0.5\}$. We repeated this
procedure 1000 times.
%We are also interested in measuring the robustness of our methodology
%to the distance used in the hierarchical clustering step, consequently, we utilized \texttt{phenopar} 
%varying the argument \texttt{distance=("dtw\_basic", "dtw2")}. Also,
%we utilized the argument \texttt{numFreq=1} to fit a harmonic regression model with one
%frequency. 
In order to assess our results we employed the mean squared error (MSE) of the estimates produced by \texttt{sephora}. The results of this simulation study are summarized in the Table~\ref{tab:tabla-MSE-sims} (in its first 6 rows).

\begin{table}[htb]
\caption{\label{tab:tabla-MSE-sims}The MSE of phenological parameter estimates (as 
provided by \texttt{phenopar}) under iid error noise, based on 1000 pseudo samples 
each of size 552. Base signal is described in the Section~\ref{ssec:sims-intro}.}
\centering
\scalebox{0.75}{
\begin{tabular}{ccc|rrrr}
\toprule[1.25pt]
 $\sigma$ & $df$ & Distance & GU & SoS & Mat & EoS\\
\midrule
0.15 & & $\mbox{Basic}$ & 0.0834128 & 0.1129451 & 0.1026047 & 0.0927606\\
0.15 & & $\mbox{L2}$ & 0.0834128 & 0.1129451 & 0.1026047 & 0.0927606\\
0.25 & & $\mbox{Basic}$ & 0.0834128 & 0.1129451 & 0.1026047 & 0.0927606\\
0.25 & & $\mbox{L2}$ & 0.0834128 & 0.1129451 & 0.1026047 & 0.0927606\\
 0.5 & & $\mbox{Basic}$ & 0.1237819 & 0.1129451 & 0.1026047 & 0.0927606\\
 0.5 & & $\mbox{L2}$ & 0.1237819 & 0.1129451 & 0.1469444 & 0.1351150\\
%   1 & & $\mbox{Basic}$ & 0.1237819 & 0.1129451 & 0.1026047 & 0.0927606\\
%   1 & & $\mbox{L2}$ & 0.1237819 & 0.1592702 & 0.1469444 & 0.1351150\\
       & 1 & $\mbox{Basic}$ & 0.0833804 & 0.0745310 & 0.0662063 & 0.0927262\\
       & 1 & $\mbox{L2}$ & 0.0833804 & 0.0745310 & 0.0662063 & 0.0927262\\
\bottomrule[1.25pt]
\end{tabular}}
\end{table}

\subsection{Heteroscedastic error noise}

Some investigations suggest that the variability of an NDVI time series depends on the
date at which every image has been acquired, \cite{borgogno2016}. In order to assess the performance of our procedure in this context, we employed the signal described in the previous section
but now we perturbed it as follows. First, we added a seasonally dependent trend to each 
of the 23 seasons within the base signal. To be more precise, to each season, we summed 
up a randomly generated number distributed 
Gaussian with zero mean and standard deviation $0.7229$ (this is the standard deviation of the 
23 observations introduced in Section~\ref{ssec:sims-intro}) to each observation within a season. Next, we added a heavy-tailed heteroscedastic noise as follows. 
We produced 23 randomly generated numbers distributed $\chi^2$ with $df$ degrees of freedom,
the first element of this set of numbers is added to the first observation within each
season, the second element is added to each season's second observation and so on.
%As in the previous section, here
%we utilized the argument \texttt{numFreq=1} to fit a harmonic regression model with one
%frequency. 
The MSE of 1000 simulations of the procedure just described are shown in the
Table~\ref{tab:tabla-MSE-sims} (in its last 2 rows).

\begin{table}[hbt]
\caption{\label{tab:tabla-MSE-sims-mispec}The MSE of phenological parameter estimates (as provided by 
\texttt{phenopar}) under heteroscedastic error noise, based on 1000 pseudo samples each of size 552. Base signal 
is described in the Section~\ref{ssec:sims-intro}.}
\centering
\scalebox{0.75}{
\begin{tabular}{lccc|rrrr}
\toprule[1.25pt]
\texttt{numFreq} & $\sigma$ & $df$ & Distance & GU & SoS & Mat & EoS\\
\midrule
2 &&&&&&&\\
& 0.15 & & $\mbox{Basic}$ & 0.0099451 & 0.0533524 & 1.157987 & 0.5555487\\
& 0.15 & & $\mbox{L2}$ & 0.0099451 & 0.0533524 & 1.157987 & 0.5555487\\
& 0.25 & & $\mbox{Basic}$ & 0.0013511 & 0.1275044 & 2.303049 & 0.7598426\\
& 0.25 & & $\mbox{L2}$ & 0.0013511 & 0.1275044 & 2.303049 & 0.7598426\\
& 0.5 & & $\mbox{Basic}$ & 0.0231901 & 0.4517804 & 5.758904 & 0.9954390\\
& 0.5 & & $\mbox{L2}$ & 0.0231901 & 0.4517804 & 5.758904 & 0.9954390\\
%   1 & & $\mbox{Basic}$ & 0.1634891 & 1.1029795 & 9.180347 & 1.1251494\\
%   1 & & $\mbox{L2}$ & 0.1634891 & 1.1029795 & 9.180347 & 1.1251494\\
&       & 1 & $\mbox{Basic}$ & 0.0774567 & 0.0282363 & 1.936418 & 0.3098269\\
&       & 1 & $\mbox{L2}$ & 0.0774567 & 0.0282363 & 1.936418 & 0.3098269\\
\midrule
3 &&&&&&&\\
& 0.15 & & $\mbox{Basic}$ & 0.7147583 & 0.2757449 & 1.136317 & 1.5619293\\
& 0.15 & & $\mbox{L2}$ & 1.2044812 & 0.3458944 & 1.420886 & 0.8736701\\
& 0.25 & & $\mbox{Basic}$ & 1.9953175 & 0.7059069 & 2.465187 & 1.2628012\\
& 0.25 & & $\mbox{L2}$ & 2.5651167 & 0.3458944 & 3.094547 & 0.7598426\\
& 0.5 & & $\mbox{Basic}$ & 2.9846901 & 0.6039916 & 4.044874 & 1.2628012\\
& 0.5 & & $\mbox{L2}$ & 433.9421656 & 0.4239853 & 12.938639 & 0.3839472\\
%   1 & & $\mbox{Basic}$ & 0.1634891 & 1.1029795 & 9.180347 & 1.1251494\\
%   1 & & $\mbox{L2}$ & 0.1634891 & 1.1029795 & 9.180347 & 1.1251494\\
&       & 1 & $\mbox{Basic}$ & 2.565117 & 0.1592702 & 3.79538 & 4.910575\\
&       & 1 & $\mbox{L2}$ & 2.565117 & 0.1592702 & 3.79538 & 4.910575\\
\bottomrule[1.25pt]
\end{tabular}}
\end{table}

\subsection{Misspecified model}

In the previous studies, we assumed that the model was correctly specified, that is,
we used the correct number of frequencies (\texttt{numFreq=1}) in the harmonic regression
model that estimates the one-frequency base signal intoduced in Section~\ref{ssec:sims-intro}.
In this study, we considered the setups described in the previous two studies, but used
\texttt{numFreq=2} and \texttt{numFreq=3} in the estimation process. For the case of homoscedastic errors, when \texttt{numFreq=3} and $\sigma < 0.5$
we had to use a \texttt{basis} built with \texttt{samples=75} for the algorithm described in
Section~\ref{subsec:fpca.algo} to converge. When $\sigma=0.5$ it was necessary to utilize
$k=2$ for the estimation algorithm to converge. Table~\ref{tab:tabla-MSE-sims-mispec}
summarizes the results of this study.

\section{Applications}~\label{sec:apps}
%-------------------------------------------------------------------------

\begin{figure}[ht]
\includegraphics[width=0.65\linewidth,height=0.35\textheight]{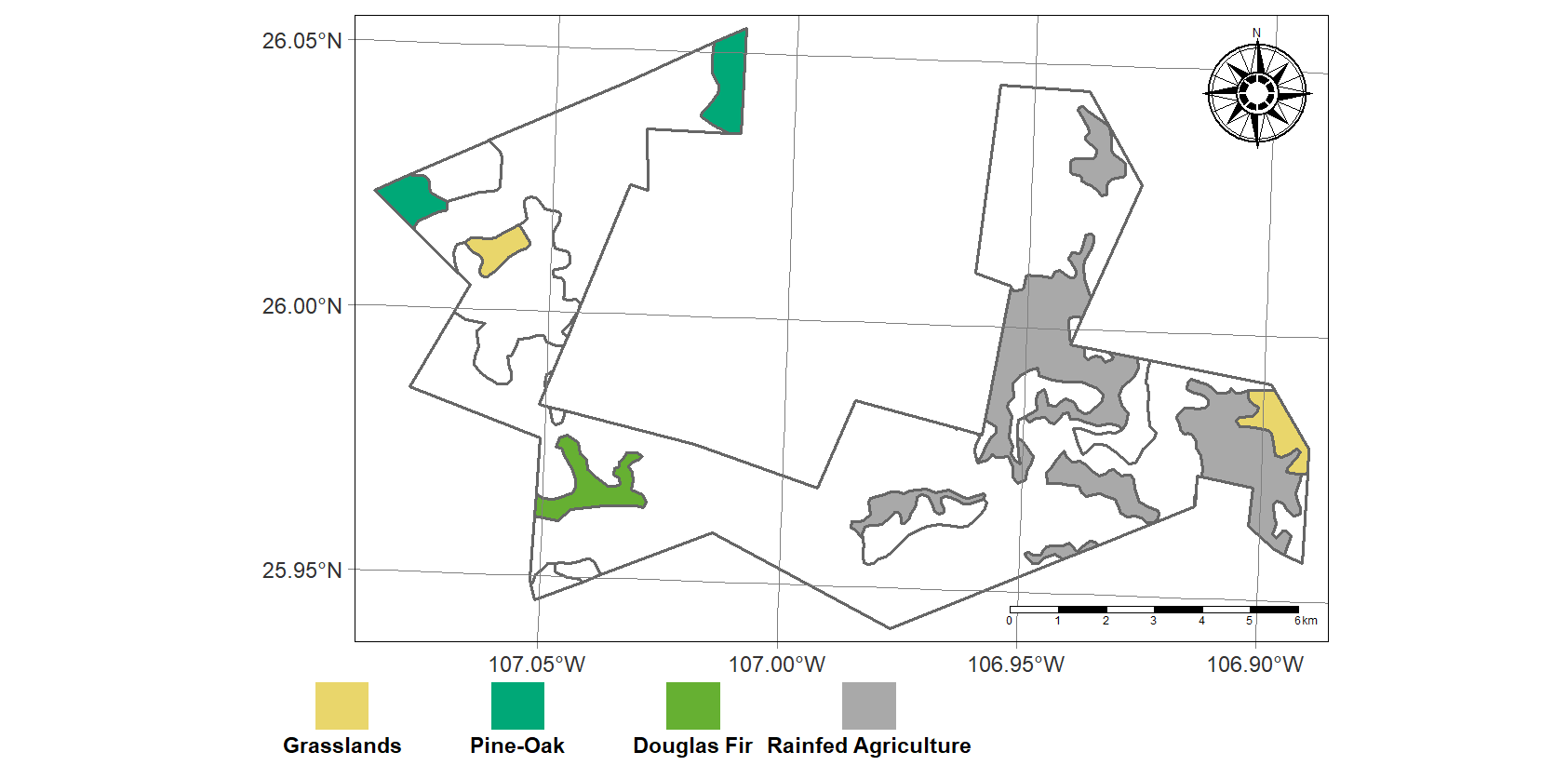} 
\caption{Flora and Fauna Protected Area Cerro Mohinora at Chihuahua, Mexico.}\label{fig:mohinora}
\end{figure}

In this section we apply our methodology to MOD13Q1 NDVI time series from
pixels located at the grasslands, Douglas Fir forest, Pine-Oak forest and
annual rainfed agriculture at the Flora and Fauna Protected Area Cerro Mohinora; see Figure \ref{fig:mohinora}. These four cover lands were generated by INEGI in
its latest national land use cartography.

\hypertarget{working-at-the-pixel-level}{%
\subsection{Working at the pixel level}\label{working-at-the-pixel-level}}

\begin{figure}[ht]
\includegraphics[width=350px, height=225px,]{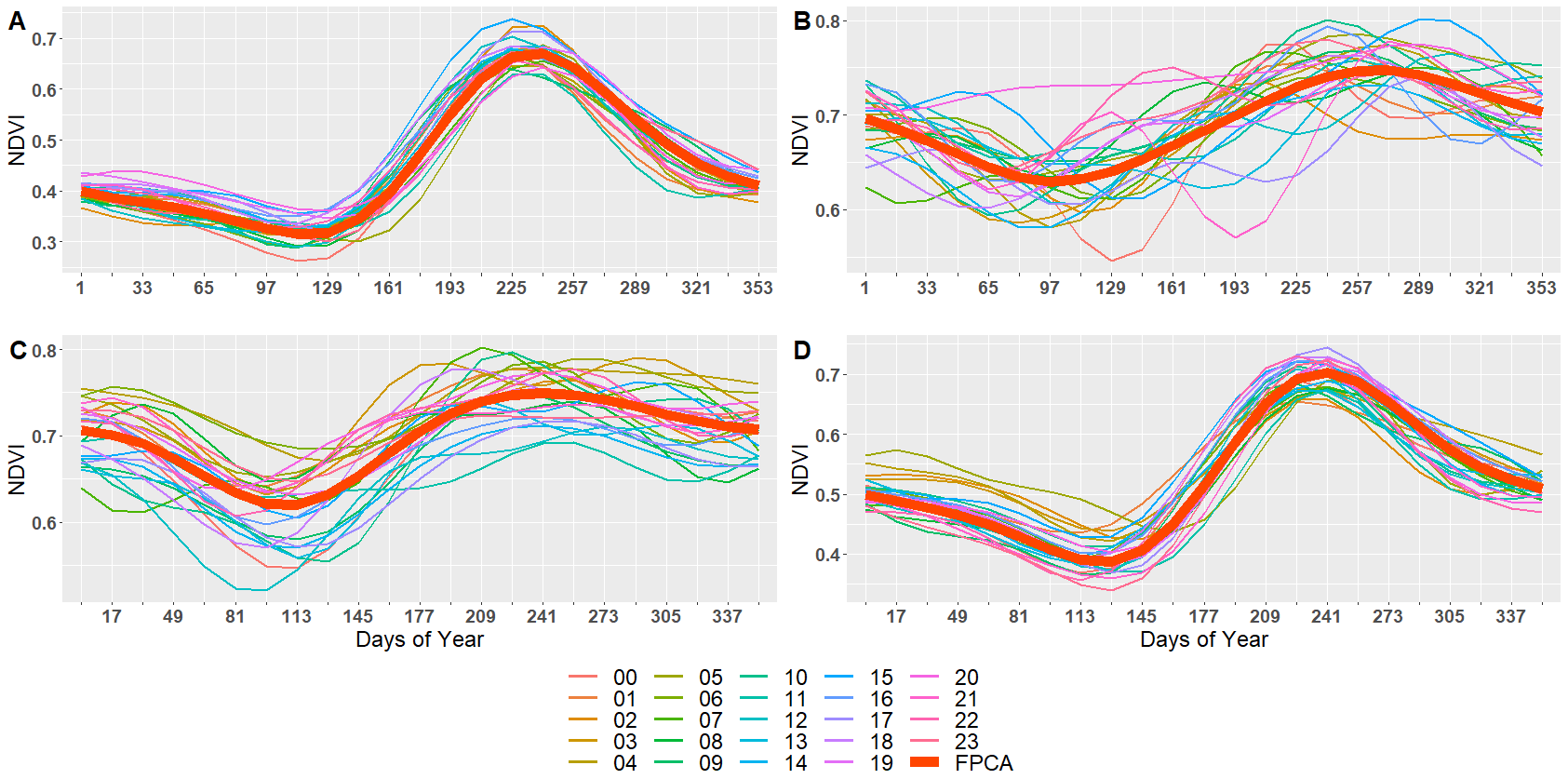}
\caption{NDVI profile plots with idealized NDVI taken from: (A) Grasslands, (B) Douglas Fir forest, (C) Pine-Oak forest and (D) Annual rainfed agriculture.}\label{fig:mohinora-pixels}
\end{figure}

In Figure \ref{fig:mohinora-pixels} we show annual, smoothed, re-sampled
and overlapped NDVI time series from some pixels located at the four cover lands
shown in Figure \ref{fig:mohinora}. We will call \emph{profile plot} to this type
of plot. For smoothing we applied a harmonic regression model with three frequencies
to each of the original, annual NDVI time series. Resampling was employed to enhance the interpretation of the estimated phenological dates which, naturally, occur at some
days along a solar year.

Typically, grasslands and rainfed agricultural areas show a quick reaction to
changes in the climatic conditions. For instance, in grasslands it is expected to observe small NDVI values (\textless{} 0.4) throughout the dry season. This characteristic
starts to change even right after the first rain of the season, starting from which we can
expect a steady sequence of increments in NDVI values until the maturity of the
vegetation is reached. Having overpassed maturity, it is expected to observe a decreasing behavior in NDVI curves. Thus, NDVI time series from pixels located at these
land covers are expected to exhibit a distinctive periodic behavior, see panels
A and D in Figure \ref{fig:mohinora-pixels}.

Unlike grasslands and agricultural areas, NDVI recorded at forests tend to show a less clear periodic
annual behavior. In panels B and C of Figure \ref{fig:mohinora-pixels} we include
profile plots of pixels located at the Douglas Fir and Pine-Oak forests of
Cerro Mohinora. As shown in these figures, throughout the year, NDVI values are
rather high, (for both sets of time series, the median and standard deviation are close to 0.7 and
0.05, respectively), in contrast with NDVI values registered at grasslands (the set of
time series of panel A has median 0.41 and s.d. 0.11) and agricultural areas (the set
of time series of panel D has median 0.51 and s.d. 0.10).

We applied a hierarchical clustering to each of the four groups of smoothed and
re-sampled NDVI time series of Figure \ref{fig:mohinora-pixels}. We used
the basic dynamic time warping distance provided by package X in each case;
although similar results were obtained with other distances.
We only allowed for the creation of two clusters.
For the pixel shown in panel A the smaller cluster was comprised by the years 2015, 2016, 2017, 2019 and 2020 allocating the remaining 19 years to the larger cluster. We found a similar behavior in the pixel of the panel D, the smaller cluster was formed by the years 2001-2005, though.
For the pixel of the panel B, located at the Douglas Fir forest, hierarchical clustering suggested to discard the year 2020 leaving the dominating cluster with 23 years. Finally, for
the pixel in the panel C the resulting clusters have sizes 13 and 11.

In addition to the profile plots, Figure \ref{fig:mohinora-pixels}'s panels
include the \emph{idealized} NDVI (thick, red line) produced by our methodology.
Each idealized NDVI is the estimated trend curve of model (X) based
on the smoothed, re-sampled NDVI curves in the \emph{dominating} cluster obtained via
the aforementioned hierarchical clustering. We defined a dominating cluster as that
cluster with at least 15 curves -which represents slightly
more than the 60\% of the total amount of curves in each group.
Hence, in Figure \ref{fig:mohinora-pixels}
the only group without a dominating group is that in panel C, in this case we opted
for estimating the idealized NDVI based on the entire 24 annual time series.

Table \ref{tab:tabla-phenopar-examples} shows the estimates of the phenological
parameters, green up (GU), start of season (SoS), maturity (Mat), senescense (Sen),
end of season (EoS) and dormancy (Dor) for each of the four groups of time series
discussed so far. These estimates are the critical points, see Eq. (W), of the idealized NDVI curve of each time series set.

\begin{table}[t]
\begin{tabular}{l|rlrlrlrl}
\toprule
%\multicolumn{1}{c}{ } & \multicolumn{2}{c}{Grasslands} & \multicolumn{2}{c}{Douglas Fri} & \multicolumn{2}{c}{Pine-Oak} & \multicolumn{2}{c}{Rainfed agriculture} \\
%\cmidrule(l{3pt}r{3pt}){2-3} \cmidrule(l{3pt}r{3pt}){4-5} \cmidrule(l{3pt}r{3pt}){6-7} \cmidrule(l{3pt}r{3pt}){8-9}
  & DoY & Date & DoY & Date & DoY & Date & DoY & Date\\
\midrule
GU & 144 & May 24 & 97 & April 7 & 109 & April 19 & 144 & May 24\\
SoS & 183 & July 2 & 184 & July 3 & 157 & June 6 & 185 & July 4\\
Mat & 228 & Au. 16 & 262 & Sep. 19 & 31 & Jan. 31 & 229 & Au. 17\\
Sen & 66 & March 7 & 20 & Jan. 20 & 196 & July 15 & 63 & March 4\\
EoS & 280 & Oc. 7 & 46 & Feb. 15 & 63 & March 4 & 282 & Oc. 9\\
%\addlinespace
Dor & 312 & No. 8 & 333 & No. 29 & 329 & No. 25 & 313 & No. 9\\
\bottomrule
\end{tabular}
\caption{\label{tab:tabla-phenopar-examples}Estimated phenological dates (in days of the year, 
DoY, and regular dates) for the pixels shown in Figure \ref{fig:spiralPlot}.}
\end{table}

There exists an evident similarity in the estimated
phenological cycles of grasslands and rainfed agricultural pixels. In both cases,
the green up occurs at the same date, their seasons start and end within a difference of two days, there is a difference of one day in their corresponding maturity and
dormancy points. When studying vegetation's dynamics, it is relevant to know
the length of the season, that is, the difference between the end and start of the
season dates. Based on our procedure, the season's length is 97 days for the
pixels shown in panels A and D of Figure \ref{fig:mohinora-pixels}.

Based on Table \ref{tab:tabla-phenopar-examples}, our procedure is able to
estimate at least three out of the six phenological parameters that comprise the
entire cycle for NDVI recorded at forests. Indeed, for the Douglas Fir forest
pixel, the estimated dates for green up, start of season, maturity and
dormancy follow a sensible chronological order. For the Pine-Oak forest pixel,
whose dynamics seems to be more intricated, our approach produces meaningful estimates
for green up, start of season and dormancy.

Unfortunately, the senescence's estimate produced by our approach does not seem
to be realistic.

\hypertarget{working-at-the-polygon-level}{%
\subsection{Working at the polygon level}\label{working-at-the-polygon-level}}

The analysis shown in the previous section provides evidence that our approach
to estimate the phenological cycle will produce better results on land covers with
a vegetation exhibiting a clear periodic dynamics such as grasslands and annual
rainfed agricultural zones. In this section we present the results of applying
our approach to all the pixels of each of the polygons shown in Figure \ref{fig:mohinora}.

The grasslands, Douglas Fir forest, Pine-Oak forest and annual rainfed agriculture
polygons contain 62, 56, 72 and 441 pixels, respectively. The time series of each
pixel of these polygons were analyzed utilizing the same specifications discussed
in the previous section. Computation was efficiently carried out through parallel
computing via the \texttt{phenopar\_polygon} function of our R package \texttt{sephora}.

\begin{figure}
\subfloat[Grasslands\label{fig:spiralPlot-1}]{\includegraphics[width=0.5\linewidth]{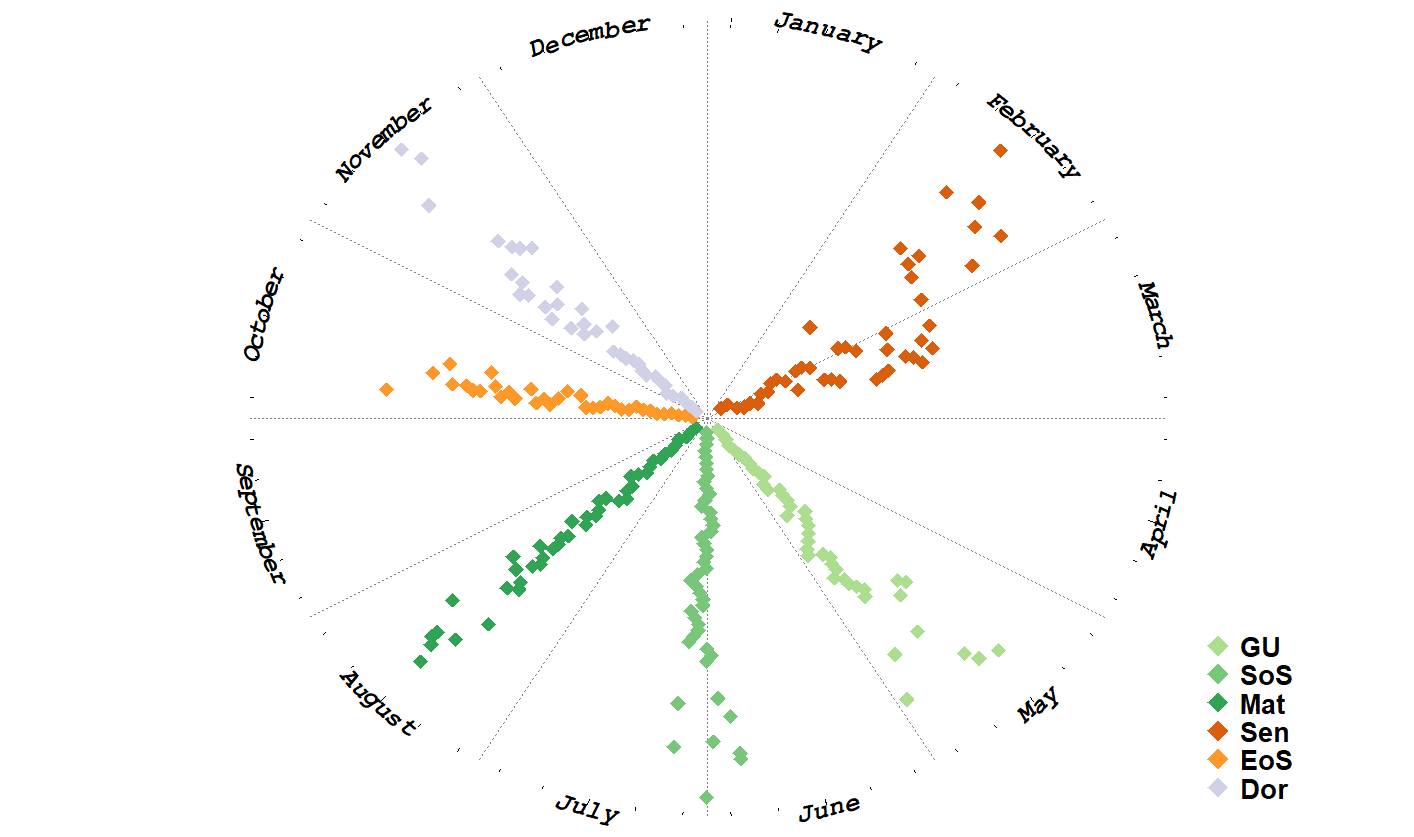} }\subfloat[Douglas Fir\label{fig:spiralPlot-2}]{\includegraphics[width=0.5\linewidth]{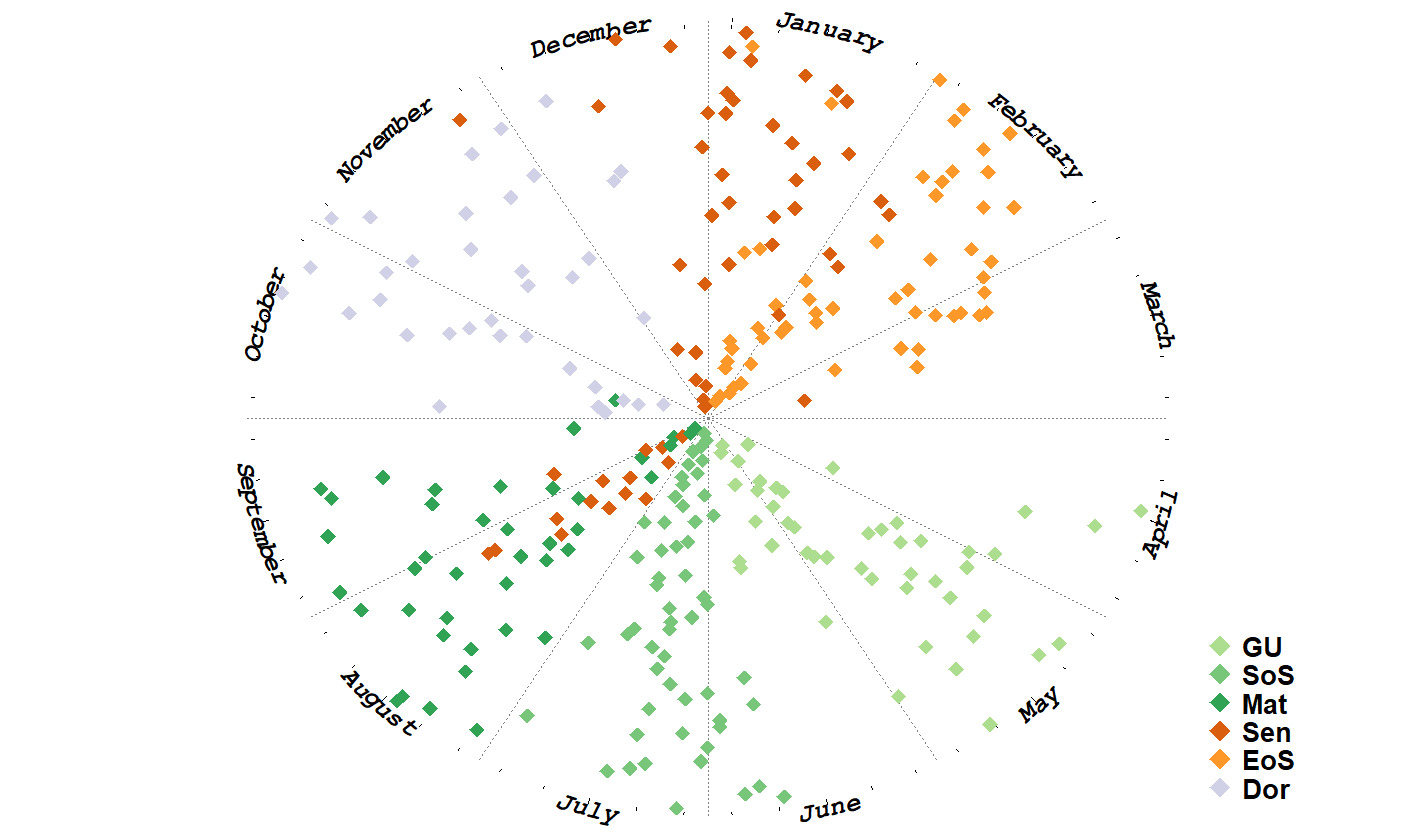} }\newline\subfloat[Pine-Oak\label{fig:spiralPlot-3}]{\includegraphics[width=0.5\linewidth]{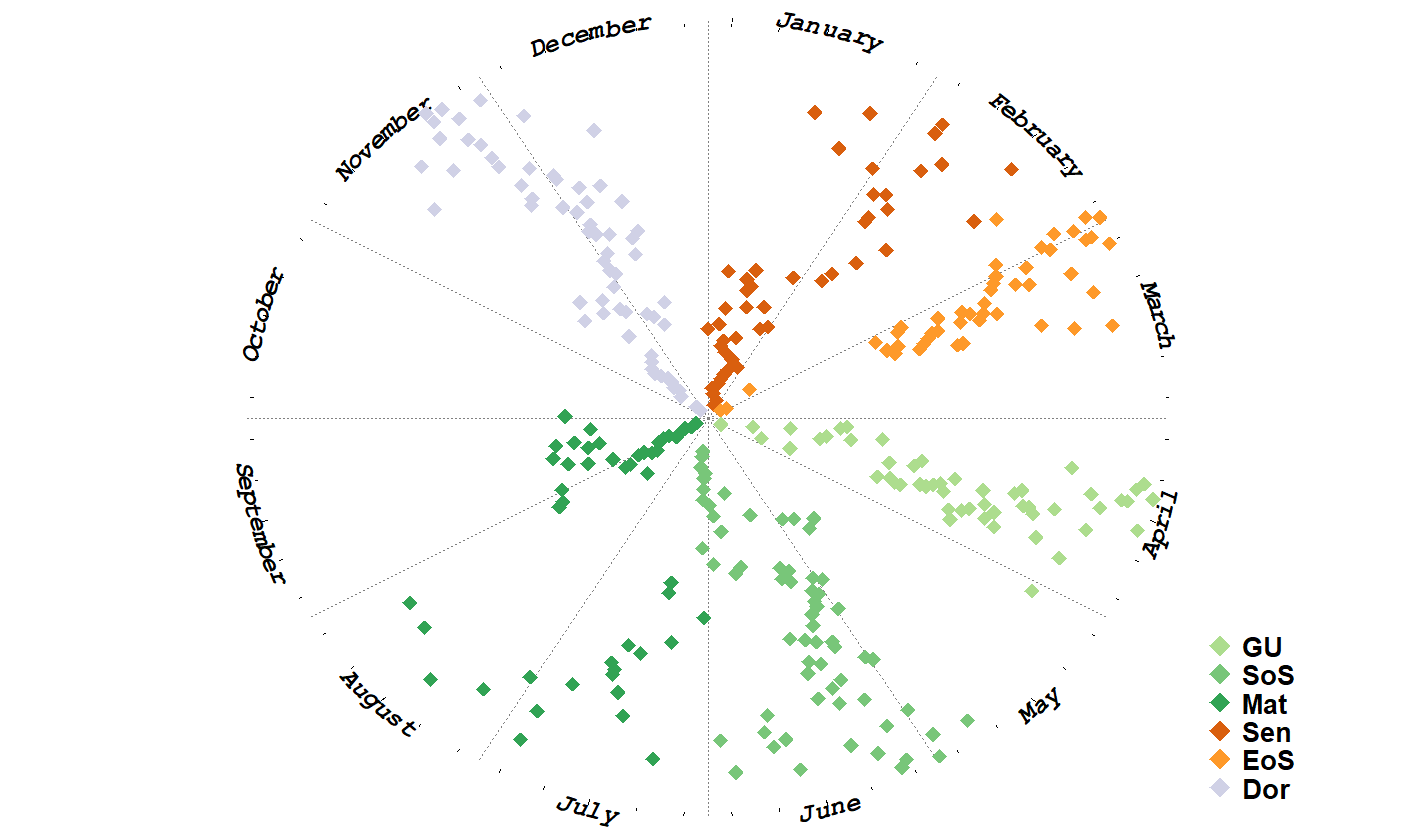} }\subfloat[Annual rainfed agriculture\label{fig:spiralPlot-4}]{\includegraphics[width=0.5\linewidth]{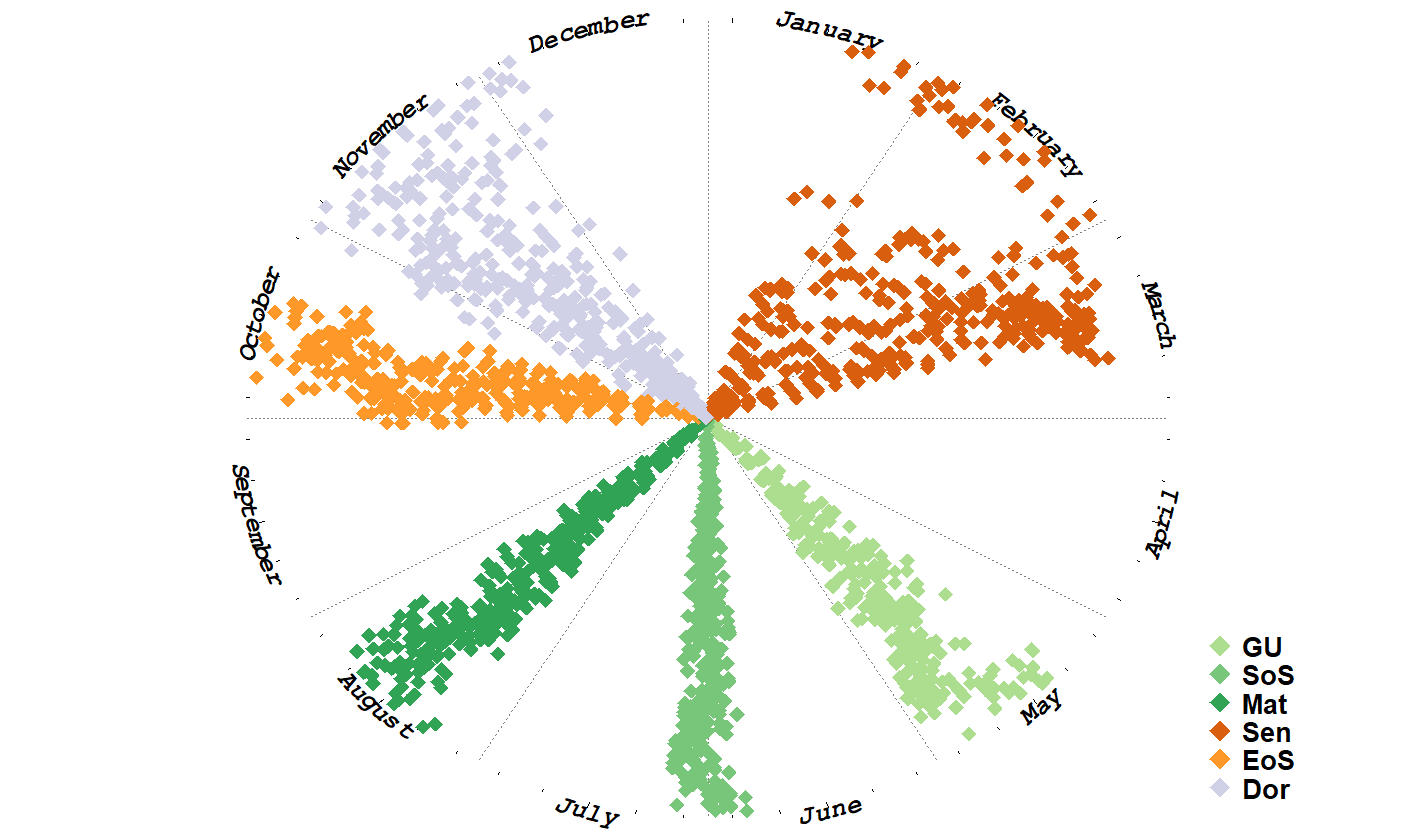} }\caption{Spiral plots of estimated phenological parameters for all the pixel in each of the polygons shown in Figure \ref{fig:mohinora-pixels}.}\label{fig:spiralPlot}
\end{figure}

After disregarding outliers and estimates lying outside an empirical 95\% confidence
interval (based on median and median absolute deviation), Figure \ref{fig:spiralPlot} shows Archimedean spirals plots displaying the estimated phenological parameters from each pixel in loops around the interval \([0, 2\pi]\);
a loop is completed once the six parameters of a single pixel are drawn in the plot.

Outliers removal varied between parameters and polygons. For grasslands, start of season (12\%) and end of season (37\%) suffered from a minimal and maximal outlier removal. In the star of season, maturity, senescence and dormancy estimates over
the Douglas Fir forest there were no outlier found and the maximum outlier
removal was observed in green up (5.35\%). In the star of season, maturity, senescence and end of season estimates over the Pine-Oak forest there were no outlier found and the maximum outlier removal was observed in green up (15.27\%). Finally, for the
Annual rainfed agriculture pixels, senescence (3.4\%) and end of season (24.7\%)
experienced a minimal and maximal outlier removal.

The main objective of this study is to characterize vegetation's phenological cycle
based on information registered in pixels from well-defined polygons. Thus,
Table \ref{tab:tabla-phenopar-trimmed} summarizes the median and median absolute
deviation (in regular dates) of each phenological parameter estimates distribution
obtained from the pixels in the polygons of Cerro Mohinora. As argued above, disregarding the
senescence date, our approach produces sensible results about the phenological
cycle in vegetations with pseudo-periodic seasons such as grasslands
and annual rainfed agriculture.

\begin{table}[t]
\begin{tabular}{l|llll}
\toprule
  & Grasslands & Douglas Fir & Pine-Oak & Agriculture\\
\midrule
GU & May 24 ($\pm$ 3 days) & May 21 ($\pm$ 16 days) & April 20 ($\pm$ 7 days) & May 23 ($\pm$ 4 days)\\
SoS & July 2 ($\pm$ 3 days) & July 11 ($\pm$ 9 days) & June 10 ($\pm$ 13 days) & July 2 ($\pm$ 3 days)\\
Mat & Au. 17 ($\pm$ 1 days) & Au. 25 ($\pm$ 13 days) & Au. 27 ($\pm$ 24 days) & Au. 17 ($\pm$ 4 days)\\
Sen & March 4 ($\pm$ 13 days) & Au. 7 ($\pm$ 215 days) & Jan. 24 ($\pm$ 15 days) & March 2 ($\pm$ 19 days)\\
EoS & Oc. 9 ($\pm$ 1 days) & Feb. 11 ($\pm$ 19 days) & March 7 ($\pm$ 5 days) & Oc. 11 ($\pm$ 6 days)\\
Dor & No. 12 ($\pm$ 3 days) & Oc. 29 ($\pm$ 19 days) & No. 26 ($\pm$ 9 days) & No. 14 ($\pm$ 10 days)\\
\bottomrule
\end{tabular}
\caption{\label{tab:tabla-phenopar-trimmed}Median and median absolute deviation of estimated phenological dates for the polygons shown in Figure \ref{fig:mohinora-pixels}.}
\end{table}

%The main objective of this study is to characterize vegetation's phenological cycle
%based on information registered in pixels from well-defined polygons. Thus,
%Table \ref{tab:tabla-phenopar-trimmed} summarizes the median and median absolute
%deviation (in regular dates) of the distribution of phenological parameter estimates
%obtained in each polygon of Cerro Mohinora. As argued above, disregarding the
%senescence date, our approach produces sensible results about the phenological
%cycle in vegetations with pseudo-periodic seasons such as grasslands
%and annual rainfed agriculture.

%\end{document}

% ------------------------------------------------------------------------

% ------------------------------------------------------------------------
\section{Acknowledgments}
%-------------------------------------------------------------------------
We are grateful to Lizbeth Naranjo-Albarr\'an and Ruth Fuentes-Garc\'ia
with Facultad de Ciencias at Universidad Nacional Aut\'onoma de M\'exico
for helpful discussions that contributed to improve this work. This paper
also benefited from discussions with Leticia Ram\'irez at CIMAT, Guanajuato.
Colors in the spiral plots were kindly suggested by Daniela Jurado with
CONABIO.
% ------------------------------------------------------------------------

% ------------------------------------------------------------------------
%\section*{Supplementary Materials for ``Phenology curve estimation via functional principal component analysis: Characterizing time series of a satellite-based vegetation index'' by
%Inder Tecuapetla-G\'omez, Francisco Rosales-Marticorena and Fanny Galicia-G\'omez}
% ------------------------------------------------------------------------

\appendix

% ------------------------------------------------------------------------
\section[Appendix A\hfill]{Proofs of claims used in Section~\ref{sec:methods}}~\label{sec:appendix}
% ------------------------------------------------------------------------

\begin{pf}[Proof of Proposition~\ref{prop:singleHarmonics}]
It is immediate 
that
\begin{align}
  f^{(1)}(t) 
  &=
  -c_1\,\left(\frac{2\pi}{L}\right)\,\sin\left( \frac{2 \pi t}{L} -\varphi_1 \right) \notag \\
  f^{(2)}(t)
  &=
  -c_1\,\left(\frac{2\pi}{L}\right)^2\,\cos\left( \frac{2 \pi t}{L} -\varphi_1 \right) \notag \\
  f^{(3)}(t)
  &=
  -\left(\frac{2\pi}{L}\right)^2\,f^{(1)}(t) \label{eq:HR-1fr-thrDerivative}\\
  f^{(4)}(t)
  &=
  -\left(\frac{2\pi}{L}\right)^2\,f^{(2)}(t). \label{eq:HR-1fr-fourthDerivative}
\end{align}
\end{pf}

\bibliographystyle{apalike}
\bibliography{sephoraBib}  % name your BibTeX data base

%
% ------------------------------------------------------------------------
\end{document}